\documentclass[journal] {IEEEtran}

\usepackage{epsfig}
\usepackage{amsmath}
\usepackage{subfigure}

\begin{document}



\title{Modeling and Analysis of Composite Antenna Superstrates Consisting on Grids of Loaded Wires}

\author{Pekka~M.~T.~Ikonen,~\IEEEmembership{Student Member,~IEEE,}
        Elena Saenz,~\IEEEmembership{Student Member,~IEEE},\\
        Ram\'on Gonzalo,~\IEEEmembership{Member,~IEEE}, and
        Sergei~A.~Tretyakov,~\IEEEmembership{Senior Member,~IEEE}
\thanks{P.~M.~T.~Ikonen and S.~A.~Tretyakov are with the Radio Laboratory/SMARAD Centre of Excellence,
Helsinki University of Technology, P.O. Box 3000, FI-02015 TKK, Finland. (e-mail: pekka.ikonen@tkk.fi). E.~Saenz
and R.~Gonzalo are with the Department of Electrical and Electronic Engineering, Public University of Navarra,
Campus Arrosadia E-31006, Pamplona-Navarra, Spain.}}

\maketitle

\begin{abstract}

We study the characteristics and radiation mechanism of antenna superstrates based on closely located periodical
grids of loaded wires. An explicit analytical method based on the local field approach is used to study the
reflection and transmission properties of such superstrates. It is shown that as a result of proper impedance
loading there exists a rather wide frequency band over which currents induced to the grids cancel each other,
leading to a wide transmission maximum. In this regime radiation is produced by the magnetic dipole moments
created by circulating out-of-phase currents flowing in the grids. An impedance matrix representation is derived
for the superstrates, and the analytical results are validated using full-wave simulations. As a practical
application example we study numerically the radiation characteristics of dipole antennas illuminating
finite-size superstrates.

\end{abstract}


\section{Introduction}

Artificial dielectrics, implemented e.g.~as arrays of wires, have been used for decades as light-weight beam
shaping elements, see e.g.~\cite{Kock_IRE}--\cite{Bahl1} for some early contributions and \cite{Pekka1, Pekka2}
for some more recent results. It is nowadays well understood that some electromagnetic (photonic) crystals
simulate the behavior of homogeneous materials with ultra-low refractive index near stop-band edges
\cite{Gralak, Notomi, Schwartz}. This observation has lead to several studies concerning directive antenna
design using such periodical structures to shape the beam of a low-gain primary radiator,
e.g.~\cite{Enoch_PRL}--\cite{Lovat}. It is also known that a slight local change in the period of
electromagnetic crystals  leads to localized resonant modes which can be used for the realization of devices
radiating energy in a very narrow angular range \cite{Tayeb}. Typically, radiating sources are placed inside
Fabry-Perot resonant cavities formed by removal of rows of wires or dielectric rods in a periodical lattice
\cite{Thevenot1}--\cite{Mittra_IEEE}. The long-known ideas of creating resonant cavities using partially
reflecting sheets \cite{Trentini} have also been widely employed in recent gain enhancement designs
\cite{Feresidis_IEE}--\cite{Boutayeb_JEWA}. Similarly, frequency-selective surfaces on top of ground planes have
been used to implement Fabry-Perot resonant cavities, e.g.~\cite{Mittra_MOTL}. Antenna superstrates are another
well-known gain enhancement tool \cite{Alexopoulos1, Alexopoulos2}. As in the case of defected electromagnetic
crystal structures \cite{Thevenot1}--\cite{Mittra_IEEE} one of the key ideas of different superstrates is to
allow radiation from a primary source to spread over a larger radiating aperture. Recently, very exotic
structures such as material covers or superstrates aimed to possess double-negative behavior \cite{Smith_meta}
have been proposed for this purpose \cite{Ziolkowski, Wu, Elena}.

The goal of this work is to present an analytical methodology for studying the characteristics of a class of
composite structures proposed as superstrates for low-gain antennas. More precisely, we consider two closely
located parallel grids of capacitively loaded wires having possibly a grid of continuous (purely inductive)
wires in between them. The methodology presented here can, however, be extended to study the characteristics of
other kinds of superstrates based on multiple grids of wires loaded with arbitrary distributed impedance. The
physical mechanism responsible for the operation of the proposed superstrates is clarified. We also briefly
discuss the extension of the methodology to handle strongly coupled dipole grids. The transmission properties of
such grids have in the past been analyzed with an application to frequency-selective surfaces or electromagnetic
band-gap structures, e.g.,~in \cite{Var1, Var2, Fer}. In these works the response has been calculated using an
iterative method based on solving a set of coupled integral equations for the unknown element currents and
aperture fields. The method presented here is explicit, and despite taking into account all the near field
interactions between the grids (higher-order Floquet modes) it remains computationally efficient and physically
illustrative.

The novelty of the proposed superstrate designs is to excite resonant oscillations in capacitively loaded wires
(this way creating a wider radiating aperture) while maintaining partial transparency of the structure. In the
designs considered here the source field induces out-of-phase currents to the loaded wire pairs, and the pairing
currents give rise to resonant magnetic dipole moments that produce aperture radiation. After heuristically
explaining the operational principle of the proposed superstrates we move on calculating the reflection and
transmission properties of the superstrates. The analytical results are validated using full-wave simulations,
and for a physical illustration we calculate the electric and magnetic dipole moments induced to one of the
structures. It is shown that strong resonant enhancement of the superstrate aperture field can be achieved with
very thin structures. To further clarify the physics behind the superstrate operation we derive an impedance
matrix representation directly from the electrodynamic model. Finally, a high-gain antenna structure utilizing
the proposed superstrates is implemented in a commercial method-of-moments solver.

\section{Heuristic operational principle}

\begin{figure}[b!]
\centering \epsfig{file=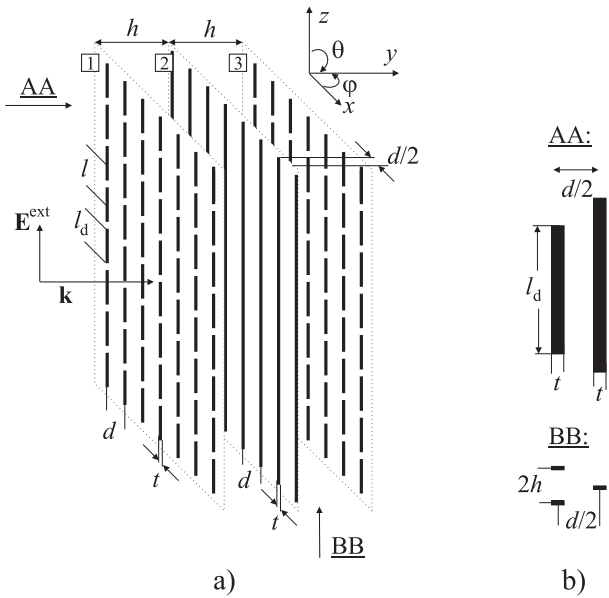, width=8.0cm} \caption{a) A schematic illustration of the proposed superstrate
(not necessarily in scale). b) Unit cell geometry.} \label{slab}
\end{figure}

The superstrate studied in this work is schematically depicted in Fig.~\ref{slab}a. The first and third grids
(see the numbering of grids in Fig.~\ref{slab}a) consist of parallel capacitively loaded\footnote{In this work
capacitive loading is created by cutting the wires.} wires, while the second grid consists of continuous
(inductive) thin wires. Please note that in Fig.~\ref{slab} round wires have been replaced by strips to reflect
one feasible implementation technique. However, in the following analytical derivation we assume round wires
since electrically thin wires and strips behave in a similar manner. To account for the difference in the
geometry an effective radius of wires is introduced in the analytical analysis.

The properly designed structure is expected to show an analogy with double-negative behavior: In a certain
frequency range the metal strips constituting a capacitively loaded wire can be interpreted as short dipoles,
and every pair of such dipoles (sitting in grids 1 and 3) shows a magnetic resonant mode which can be utilized
to create a response similar to that of split rings \cite{Pendry, Dolling}. On the other hand, the long
conductors provide plasma-like electric response as in wire media \cite{Pavel,Maslovski}. Recently, it has been
reported that only short wire pairs are enough to obtain double-negative behavior \cite{Shalaev, Zhou}. For
describing the radiation properties of the superstrate we concentrate, however, on the distributed dipole
moments induced in the structure.

Consider first only a pair of two capacitively loaded wires (assuming that grid 2 is absent). At a certain
frequency the currents induced to individual pairing metal strips cancel each other allowing the incident wave
to propagate through the grids. To widen the pass-band one can bring a grid of solid wires in between the loaded
wire grids. In this case two modes appear due to the capacitances and mutual coupling between the loaded wires
and the continuous wires. In other words, there are now two possibilities for the currents in the grids to
cancel each other, and by properly choosing the unit cell dimensions both transmission bands can be merged
together. Since the superstrate is \emph{volumetric} (though the total thickness $2h$ is in practise extremely
small compared with the wavelength) the cancelation of total electric current does not mean the cancelation of
radiation: The amplitude of the reflection coefficient is minimized at this frequency (and consequently the
amplitude of the transmission coefficient is maximized), however, the magnetic dipole moments created by the
pairing out-of-phase
currents produce radiation. 

Another important practical issue, discussed below in more detail, is the interaction between a closely located
primary radiator and the loaded wire grids. It will be shown that in case of capacitively loaded wires the
source can be brought very close to the grids without deteriorating the radiation resistance. This allows us to
design low-profile and effectively radiating structures.

\section{Local field method to study the superstrate properties}

\subsection{Plane-wave reflection and transmission coefficients and induced electric and magnetic dipole moments}

At this point we assume that the grids in the superstrate are infinite in the $z$ and $x$ directions. Moreover,
the transversal thickness $t$ of the inclusions (equivalently the radius $r_0$ of round wires) is small compared
to the wavelength. We assume the normal plane-wave incidence (though the analysis can be carried out for oblique
incidence as well) and write the external electric field as
\begin{equation}
\textbf{E}^{\rm ext} = \textbf{u}_zE^{\rm ext}e^{-jky}.
\end{equation}
We write down the equation for the local electric field acting on the surface of a reference element in each of
the grids. Consider first the unloaded reference wire, and assume that the wires are thin enough so that we can
represent them as infinitely long current lines located at the wire's axes. The relation between the amplitude
of the local electric field $E^{\rm loc}_z$ acting on the wire surface and current $I$ flowing on the wire
surface reads
\begin{equation}
E^{\rm loc}_z = \alpha_0^{-1}I,
\end{equation}
where $\alpha_0$ is the susceptibility of the wire which can be determined from the boundary condition on the
wire surface (see e.g.~\cite{Felsen}):
\begin{equation}
\alpha_0 = \bigg{[}\frac{\eta k}{4}H_0^{(2)}(kr_0)\bigg{]}^{-1}, \label{alpha_0}
\end{equation}
where $\eta$ is the wave impedance in the matrix material (host material for the wires), and $H_0^{(2)}$
represents the Hankel function of the second kind and zero order. If the wires are loaded with a certain
uniformly distributed impedance per unit length $Z$, the inverse of the wire susceptibility becomes
\cite{Pavel_PRE, Sergei}:
\begin{equation}
\alpha^{-1} = \alpha_0^{-1} + Z.
\end{equation}
The amplitude of the local electric field acting on the reference wires can also be expressed in the following
manner:
\begin{equation}
\alpha^{-1}I_1 = E^{\rm ext} + \beta(0)I_1 + \beta(h)I_2 + \beta(2h)I_3, \label{e1}
\end{equation}
\begin{equation}
\alpha_0^{-1}I_2 = E^{\rm ext} e^{-jkh} + \beta(h)I_1  + \beta(0)I_2 + \beta(h)I_3, \label{e2}
\end{equation}
\begin{equation}
\alpha^{-1}I_3 = E^{\rm ext} e^{-j2kh} + \beta(2h)I_1 + \beta(h)I_2 + \beta(0)I_3, \label{e3}
\end{equation}
where $I_i$ ($i=1,2,3$) is the amplitude of current induced to the grid $i$, and $\beta(0)$ is so called self
interaction coefficient (eq.~(11) in \cite{Pavel}, see also \cite{Yatsenko}):
\begin{multline}
\beta(0) = -\frac{\eta k}{2}\bigg{[}\frac{1}{kd}
- \frac{1}{2} +\frac{j}{\pi}\bigg{(}\log\frac{kd}{4\pi} + \gamma\bigg{)}\\
+\frac{j}{d}\sum_{n\neq0}\bigg{(}\frac{1}{\sqrt{(2\pi n/d)^2 - k^2}} - \frac{d}{2\pi |n|}\bigg{)}\bigg{]},
\label{b0}
\end{multline}
and $\beta(h), \beta(2h)$ are mutual interaction coefficients (eqs.~(12) and (13) in \cite{Pavel}):
\begin{equation}
\beta(ih) = -\frac{\eta k}{2d}\sum_{n=-\infty}^{n=+\infty}\frac{e^{-jk_x^{(n)}ih}}{k_x^{(n)}}, \quad i=1,2.
\end{equation}
Above $\gamma\approx0.5772$ is the Euler constant, and $k_x^{(n)}$ is the $x$-component of the $n$-th Floquet
mode wave vector \cite{Pavel}:
\begin{equation}
k_x^{(n)} = -j\sqrt{(2\pi n/d)^2 - k^2}, \quad \rm{Re}\{\sqrt{()}\}>0.
\end{equation}
Physically $\beta(0)$ tells how strongly the wires located in the same grid as the reference wire  influence the
local field, whereas $\beta(h),\beta(2h)$ measure the influence of the other two grids. Now we can cast
eqs.~(\ref{e1})--(\ref{e3}) in the following form:
\begin{multline}
-E^{\rm ext}\left( \begin{array}{c}
1 \\
e^{-jkh} \\
e^{-j2kh}
\end{array} \right)=\\
\left( \begin{array}{ccc}
\beta(0) - \alpha^{-1} & \beta(h) & \beta(2h) \\
\beta(h) & \beta(0) - \alpha_0^{-1} & \beta(h) \\
\beta(2h) & \beta(h) & \beta(0) - \alpha^{-1}
\end{array} \right) \left( \begin{array}{c}
I_1 \\
I_2 \\
I_3
\end{array} \right).
\end{multline}
From the above equation currents in different grids can be solved by simple inversion. The averaged current
density induced to the grids is defined as
\begin{equation}
\widehat{J}_i=\frac{I_i}{d}, \quad i=1,2,3.
\end{equation} \label{J}

In the far zone, the reflected field of one grid is a plane-wave field, and in the grid plane this plane-wave
field can be expressed as \cite{Sergei}
\begin{equation}
E^{\rm sc}_i=-\frac{\eta}{2}\widehat{J_i}. \label{E}
\end{equation}
After this we can calculate the total reflection and transmission coefficients. The reflection and transmission
coefficients referred to the plane of the first grid read:
\begin{equation}
R = \frac{\sum_{i=1}^3{E^{\rm sc}_ie^{-jkh(i-1)}}}{E^{\rm ext}}, \label{Rpw}
\end{equation}
\begin{equation}
T = 1 + \frac{\sum_{i=1}^3{E^{\rm sc}_ie^{jkh(i-1)}}}{E^{\rm ext}}. \label{Tpw}
\end{equation}

As an example, we calculate the dipole moments induced to the structure consisting of two grids of capacitively
loaded wires (in this case grid 2 is absent in Fig.~\ref{slab}a). The electric and magnetic dipole moments
confined in the unit volume (Fig.~\ref{slab}b) are calculated as \cite{Jackson}
\begin{equation}
\textbf{p} = \int \textbf{u}\rho \hspace{0.15cm} dV,
\end{equation}
\begin{equation}
\textbf{m} = \mu_0 \int \textbf{r} \times \textbf{J} \hspace{0.15cm} dV,
\end{equation}
where $\rho$ is the charge density in the unit volume, and $\textbf{J}$ is the corresponding volumetric current
density. When the integrations are conducted in our geometry the dipole moments take the following form:
\begin{equation}
p = \frac{l_{\rm d}}{j\omega}(I_1 + I_3),
\end{equation}
\begin{equation}
m = \mu_0\frac{hl_{\rm d}}{2}(I_1 - I_3).
\end{equation}

\subsubsection{Discussion on the modeling approach}

It is known that at low frequencies (frequencies at which the structural dimensions are small compared to the
wavelength) the dispersion properties of lattices of capacitively loaded wires and dipole scatterers rather
closely resemble each other \cite{Pavel_PRE}. Noticeable deviation is seen only close to the frequencies where
the strip length in the loaded wires becomes resonant. At these frequencies (and at higher frequencies), for
accurate analysis one should model the first and third grids as grids of dipole scatterers. The principal ideas
of the above derivation would still be the same, however, different models for inclusion polarizabilities and
interaction constants should be used \cite{Maslovski_int}--\cite{Yatsenko_IEEE}.Thus, a similar local field
approach as introduced in the present paper can also be used to treat grids in which the unit inclusions can be
modeled as small  dipole scatterers. It will be demonstrated below that in the frequency regime where the
superstrates proposed in this paper are feasible for antenna beam shapers, the capacitively loaded wire model
leads to accurate results. It is also important to note that the method described above can be used to study the
properties of superstrates formed by an arbitrary number of grids, and the distributed impedance loading the
wires can be arbitrary.

\subsubsection{Numerical validation}

For the validation of the analytical method we choose for the superstrate the structural parameters considered
in \cite{Elena} (the notation is clear from Fig.~\ref{slab}): $l=12.11$ mm, $l_{\rm d}=10.11$ mm, $t=1$ mm,
$d=4.8$ mm, $h=1$ mm. We introduce equivalent round-cross-section wires, and use text book formulas to estimate
the distributed capacitance due to the slits in the wires in first and third grids.

\begin{figure}[b!]
\centering \epsfig{file=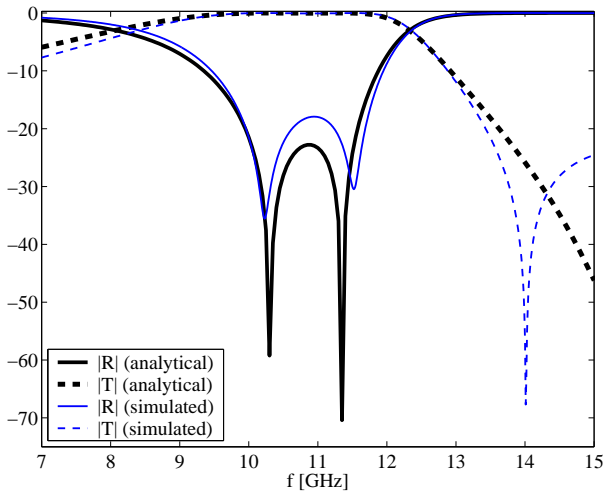, width=8cm} \caption{Calculated and simulated reflection and transmission
coefficients for two grids of capacitively loaded wires with one grid of continuous wires in between them.}
\label{c1}
\end{figure}

\begin{figure}[t!]
\centering \epsfig{file=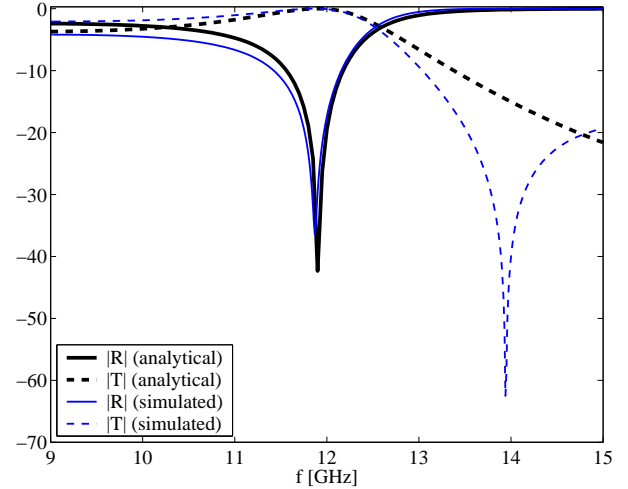, width=8cm} \caption{Calculated and simulated reflection and transmission
coefficients for two grids of capacitively loaded wires.} \label{c2}
\end{figure}

For calculating the reflection and transmission coefficients we consider separately two structures. In the first
one only the two grids of capacitively loaded wires are present. In the second case we bring the grid of
continuous wires in between the loaded wire grids as shown in Fig.~\ref{slab}a. Reflection and transmission
coefficients are calculated using eqs.~(\ref{Rpw}) and (\ref{Tpw}). We also simulate the coefficients using
Ansoft HFSS \cite{Ansoft} (the unit cell for HFSS simulations is shown in Fig.~\ref{slab}b). The results of
calculations and simulations are depicted in Figs.~\ref{c1} and \ref{c2}. We observe that the agreement between
the analytical calculations and simulations is good up to 13 GHz. At this frequency the length of the metal
strips in the loaded wires is already roughly $\lambda/2$, so it is clear that at frequencies higher than 13 GHz
(in this particular case) the local field model based on loaded wires fails. Nevertheless, for antenna
applications we are interested in the frequency regime where the superstrates are partially transparent
(frequency regime in which the reflection minima and transmission maxima lie, see Figs.~\ref{c1} and \ref{c2}),
and at these frequencies the analytical model works well.

\begin{figure}[b!]
\centering \epsfig{file=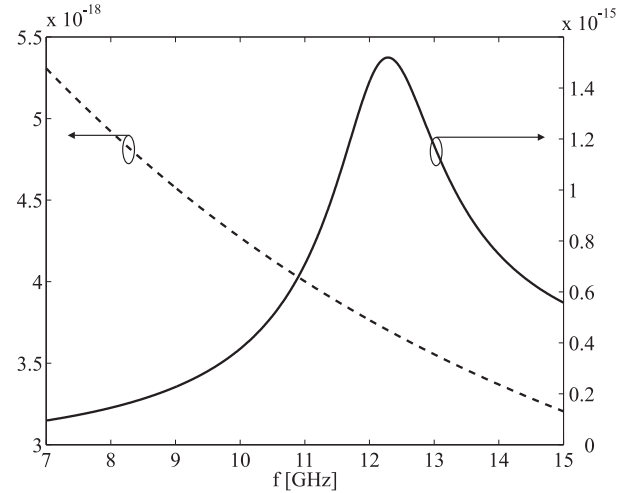, width=8cm} \caption{Dashed line:~Electric dipole moment induced in the unit
volume (in [Asm]), solid line:~Magnetic dipole moment induced in the unit volume (in [Vsm]). The superstrate
consists of two grids of capacitively loaded wires.} \label{moments}
\end{figure}

The amplitudes of electric and magnetic dipole moments induced in the unit volume of the superstrate consisting
of two grids of loaded wires are shown in Fig.~\ref{moments}. As speculated earlier, the magnetic dipole moment
shows a strong resonance in the vicinity of the frequency where the plane-wave reflection coefficient is
minimized. It is worth to note that at the frequency of magnetic resonance the thickness of the superstrate is
only $\lambda/12$.

\subsection{Impedance matrix representation}

Based on the analysis of induced currents we derive an impedance matrix representation \cite{Pozar} connecting
the total (averaged) electric field in the grid planes to the averaged current densities flowing in the grids:
\begin{equation} \left(
\begin{array}{c}
\widehat{E}_1^{\rm tot} \\
\widehat{E}_2^{\rm tot} \\
\widehat{E}_3^{\rm tot}
\end{array} \right) =
\left( \begin{array}{ccc}
Z_{11} & Z_{12} & Z_{13} \\
Z_{21} & Z_{22} & Z_{23} \\
Z_{31} & Z_{32} & Z_{33}
\end{array} \right) \left( \begin{array}{c}
\widehat{J}_1 \\
\widehat{J}_2 \\
\widehat{J}_3
\end{array} \right). \label{Z1}
\end{equation}
Such representation is useful, e.g.,~when representing the grids as equivalent discrete loads in a waveguide
modeling free space plane-wave propagation \cite{Brown}. The expressions deduced below allow also to represent
the mutual interaction between the grids at the impedance description level. In the quasi-static limit the
deduced impedance expressions can be used to construct an equivalent circuit for the superstrates. For the sake
of clarity we conduct below the derivation for the first grid in details.

Using the impedance matrix notation the total field in the plane of the first grid reads
\begin{equation}
\widehat{E}_1^{\rm tot} = Z_{11}\widehat{J}_1 + Z_{12}\widehat{J}_2 + Z_{13}\widehat{J}_3.
\end{equation}
Following the definition of the impedance matrix \cite{Pozar}, $Z_{11}$ is the total electric field in the first
grid plane divided by the averaged current flowing in that grid, assuming that currents in the other grids are
zero. The component of the total (averaged) electric field parallel to the grids, defined in the plane of grid
$i$ is by definition \cite{Sergei}
\begin{equation}
\widehat{E}^{\rm tot}_i = E^{\rm ext} + E_i^{\rm sc} = E^{\rm ext} -\frac{\eta}{2}\widehat{J}_i. \label{E_tot}
\end{equation}
When we solve $E^{\rm ext}$ from (\ref{e1}) (now $I_2=I_3=0$) and substitute the result in (\ref{E_tot}), the
following expression is obtained:
\begin{equation}
\widehat{E}^{\rm tot}_1\bigg{|}_{I_2=I_3=0} = -d\bigg{(} \beta(0) - \alpha^{-1} +
\frac{\eta}{2d}\bigg{)}\widehat{J}_1 = Z_{11}\widehat{J}_1. \label{E1_tot}
\end{equation}
Next, consider the definition of $Z_{12}$. The plane-wave term disappears in (\ref{E_tot}) since now the current
flowing in the first grid is by definition zero \cite{Pozar}. Thus, we get (after we solve $E^{\rm ext}$ from
(\ref{e1}) assuming that $I_1=I_3=0$):
\begin{equation}
\widehat{E}^{\rm tot}_1\bigg{|}_{I_1=I_3=0} = -d\beta(h)\widehat{J}_2 = Z_{12}\widehat{J}_2.
\end{equation}
Similarly for $Z_{13}$:
\begin{equation}
\widehat{E}^{\rm tot}_1\bigg{|}_{I_1=I_2=0} = -d\beta(2h)\widehat{J}_3 = Z_{13}\widehat{J}_3.
\end{equation}
After repeating the above procedure for the second and third grid we get the following expressions for the
impedances:
\begin{equation}
Z_{11} = Z_{33} = -\bigg{(}\beta(0) - \alpha^{-1} + \frac{\eta}{2d}\bigg{)}d,
\end{equation}
\begin{equation}
Z_{22} = -\bigg{(}\beta(0) - \alpha_0^{-1} + \frac{\eta}{2d}\bigg{)}d,
\end{equation}
\begin{equation}
Z_{21} = Z_{12} = Z_{32} = Z_{23} = - \beta(h)d,
\end{equation}
\begin{equation}
Z_{13} = Z_{31} = - \beta(2h)d.
\end{equation}

\section{Finite size superstrates illuminated by wire antennas}

In this section we study and discuss the performance of finite-size superstrates when illuminated by practical
antennas. Throughout the rest of the paper our primary source is a resonant $\lambda/2$-dipole antenna.
Naturally, the superstrates could also be illuminated with other types of antennas, such as with microstrip
antennas, for example.

\subsection{Large superstrate with infinitely long wires}

\begin{figure}[b!]
\centering \epsfig{file=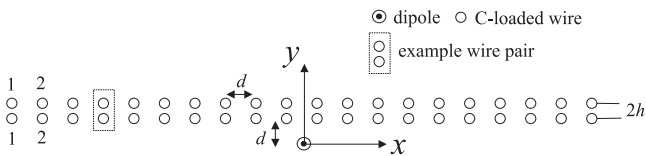, width=8.5cm} \caption{Geometry of the structure used to study the distribution
of current induced to pairing wires (not necessary in scale).} \label{asche}
\end{figure}

Before designing an actual prototype antenna we study how well the key ideas for the superstrate performance,
outlined with the plane-wave excitation, apply when the superstrates are illuminated with a resonant dipole, and
the number of wires in the superstrates is finite. Namely, we study possible cancelation of electric currents in
such structures. To avoid the standing wave interference caused by a finite wire height, and more clearly see
the possible cancelation effect, we assume in this subsection that the wires are infinitely long. To ease the
physical interpretation we consider here only the structure containing two grids of loaded wires, since in this
case there is only one possibility (one frequency) for the cancelation of the total electric current, please see
Fig.~\ref{c2}. Later, when designing the prototype antenna we consider also the superstrate containing a grid of
unloaded wires.

The number of wires in the grids is fixed to be 20, and we also fix the source position, see Fig.~\ref{asche}.
The structural parameters are the same as introduced in Section III A. To handle the infinite wire length we use
the method described in \cite{He} to solve the currents induced to the grids by taking into account the dipole
excitation and the mutual interaction between all the wires. The distribution of induced current (amplitude and
phase) is studied at three different frequencies, namely at $11.9$, $10.4$, and $13.4$ GHz. Based on the result
in Fig.~\ref{c2} we predict that 11.9 GHz corresponds to the frequency at which the amplitude of current induced
to the pairing wires is more or less the same, but the phase difference between the pairing currents is roughly
$\pi$.

\begin{figure}[t!]
\centering \epsfig{file=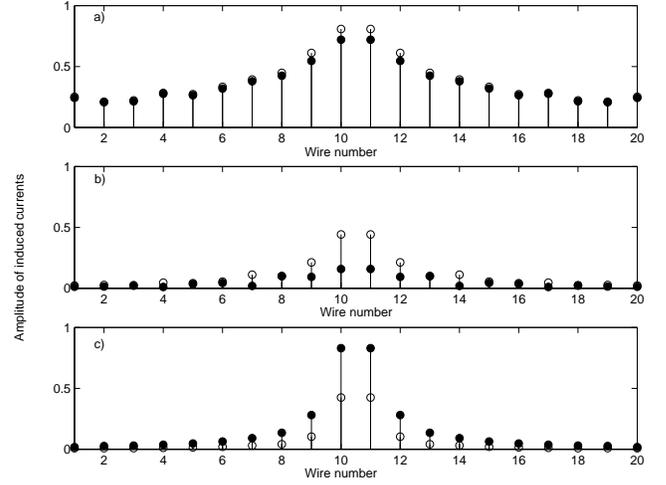, width=8.5cm} \caption{Amplitudes of currents induced to the wires at different
frequencies. Black dots denote the amplitude of current induced to wires located in the grid closest to the
dipole (Fig.~\ref{asche}), white dots denote the amplitude of current induced to the wires in the other grid. a)
$f=11.6$ GHz, b) $f=10.4$ GHz, c) $f=13.4$ GHz.} \label{amplitude}
\end{figure}

\begin{figure}[b!]
\centering \epsfig{file=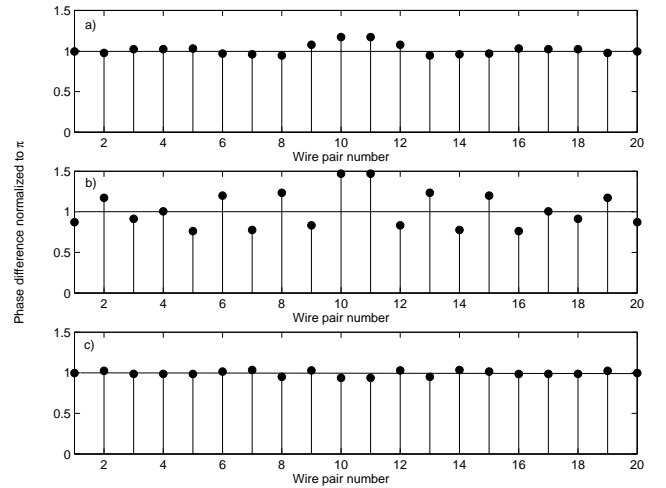, width=8.5cm} \caption{Phase difference (normalized to $\pi$) between currents
induced to the pairing wires at different frequencies. a) $f=11.6$ GHz, b) $f=10.4$ GHz, c) $f=13.4$ GHz.}
\label{phase}
\end{figure}

The amplitudes of the currents induced to the pairing wires are shown in Fig.~\ref{amplitude} (the feed current
amplitude equals unity). The phase differences of currents induced to the pairing wires are depicted in
Fig.~\ref{phase}. We observe that at 11.9 GHz the amplitudes of currents induced to the pairing wires are close
to the same value in most of the pairs. Slight difference is seen only in the wires closest to the dipole, but
this difference is compensated by the fact that also the phase difference between the currents induced to the
corresponding wire pairs slightly differs from $\pi$. In those wire pairs where the current amplitudes are
practically the same, also the phase difference is practically $\pi$. The aforementioned implies that, indeed,
the total electric current induced to the grids cancels as was speculated already in Section II. At 10.4 GHz the
amplitudes of induced currents are much lower than at 11.9 GHz. Also the amplitudes of several pairing currents
considerably differ from each other, and in many pairs the phase difference between the currents considerably
differs from $\pi$. Even though the phase differences of pairing currents is close to $\pi$ in all the pairing
wires at 13.4 GHz (Fig.~\ref{phase}c), the total electric current is not canceled at this frequency since the
amplitudes of currents constituting the pairs differ from each other in many pairs, Fig.~\ref{amplitude}c.

\subsection{High-gain antenna structure utilizing the proposed superstrates}

\begin{figure}[b!]
\centering \epsfig{file=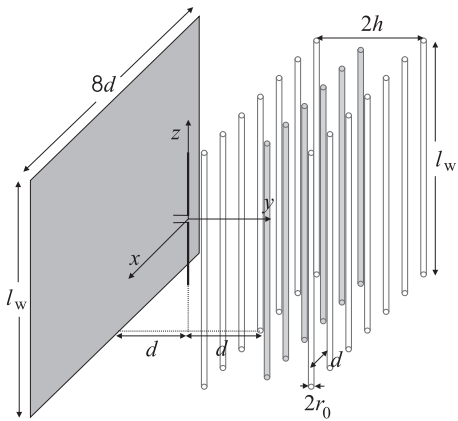, width=6cm} \caption{Schematic illustration of the prototype antenna geometry
(not necessary in scale). The non-colored rods denote capacitively loaded wires, the grey rods denote unloaded
wires. The black rod is the resonant $\lambda/2$-dipole.} \label{antenna}
\end{figure}

In this subsection we design and implement a high-gain prototype antenna in a commercial method-of-moments
solver FEKO \cite{Feko}. The schematic illustration of the studied antenna structure is depicted in
Fig.~\ref{asche}. We consider separately two structures:~In the first case the grid of unloaded wires is absent
(when analyzing the results we call this case the ``2 grids'' case), and in the second case the structure is as
depicted in Fig.~\ref{antenna} (``3 grids''). The radiation characteristics of both antenna structures are
studied at the three frequencies introduced in the previous subsection.

The dipole length is tuned to be $\lambda/2$ at every considered frequency. The wire length is fixed to be
$l_{\rm w} = 3\lambda$ at 11.9 GHz. Also the width (in $x$-direction) of the superstrate is approximately
$3\lambda$ at 11.9 GHz. To enhance the directivity of the antenna structure, and make it more attractive as a
possible high-gain solution, we create a resonant cavity using a metal reflector behind the grids,
Fig.~\ref{antenna}. Some of the available earlier contributions concerning the implementation of Fabry-Perot
resonant cavities using partially reflecting sheets or other structures are listed in the Introduction. The
distance between the metal reflector and the first wire grid equals $2d$ which roughly corresponds to
$\lambda/2$ at 11.9 GHz. The other relevant structural parameters are the same as introduced in connection with
the plane-wave analysis (see Section III A).\footnote{In FEKO we model round wires, and the capacitance caused
by slits in the strips is modeled as an equivalent bulk capacitor.} In the following the horizontal plane
(H-plane) corresponds to the plane perpendicular to the wires, and vertical (E-plane) corresponds to the plane
parallel to the wires.

\subsubsection{Example results}

\begin{figure}[b!]
\centering \epsfig{file=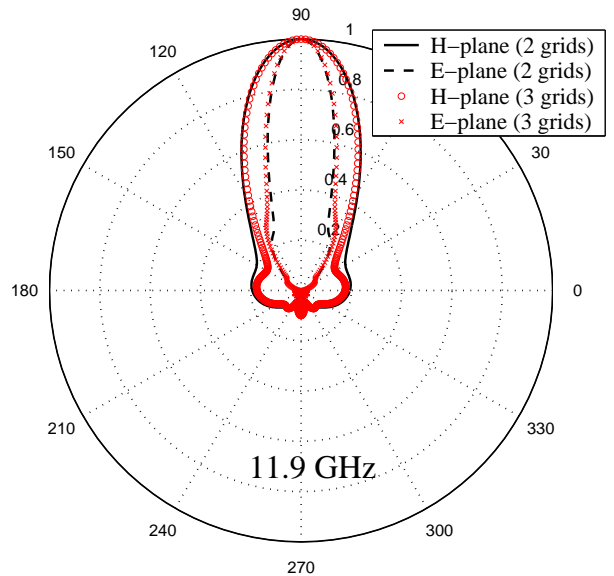, width=8cm} \caption{Radiation patterns at 11.9 GHz.} \label{119}
\end{figure}

\begin{figure}[t!]
\centering \epsfig{file=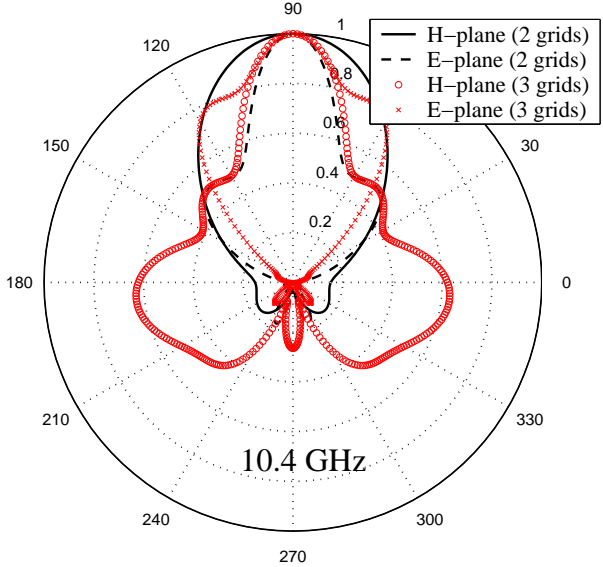, width=8cm} \caption{Radiation patterns at 10.4 GHz.} \label{104}
\end{figure}

For this study the interesting antenna characteristics are the antenna radiation pattern, the input impedance
$Z_{\rm in}$, and the maximum boresight directivity $D^{\rm max}$. We have not performed any systematic
optimization of one or more of these characteristics, thus, from many points of view the presented antenna
performance might not be optimal. The reader is also reminded that compared to the situation analyzed in Section
III where the number of wires is infinite in the $x$-direction and the wire length is infinite, the finite
number of wires and the finite wire height lead to the formation of standing waves along $x$ and $z$-directions.
This standing wave formation depends also on the location of the source, and it rather strongly influences the
radiation characteristics.

\begin{figure}[b!]
\centering \epsfig{file=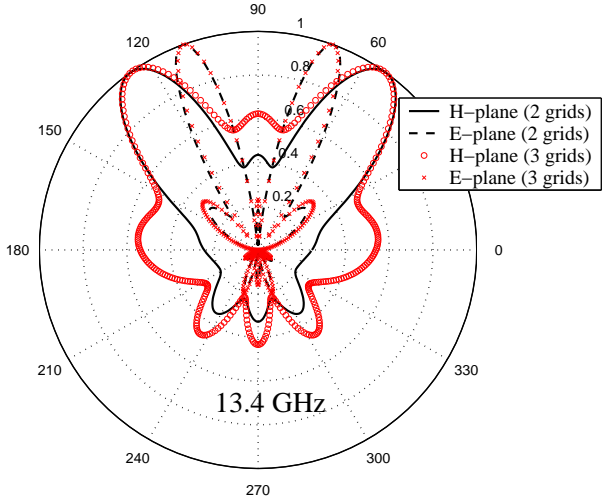, width=8cm} \caption{Radiation patterns at 13.4 GHz.} \label{134}
\end{figure}

The simulated radiation patterns are depicted in Figs.~\ref{119}--\ref{134}, and the main simulated radiation
characteristics are gathered in Table \ref{sim_table}. Even without systematically optimizing the structural
parameters we observe a good boresight directivity at 11.9 GHz, which is consistent with the analysis presented
so far in this paper. For comparison, without the superstrate the boresight directivity of a metal-plate backed
dipole is 8.5 dB. In addition to this, the input resistance (in case of lossless structures, the radiation
resistance) is high and makes the antenna easy to match to a standard 50 Ohm cable. It was analytically shown in
\cite{Sim} that when a wire antenna is positioned in close vicinity of capacitively loaded wires, the currents
induced to the closest wires add up constructively with the source current, and therefore the radiation
resistance remains high. This, indeed, appears to be the case. For comparison, in case when all the wires
depicted in Fig.~\ref{antenna} are unloaded, the simulated input resistance at 11.9 GHz is only 0.7 Ohm.

At 10.4 and 13.4 GHz the radiation characteristics are not so favorable for a feasible high-gain solution. At
10.4 GHz the peak directivity is weaker with both superstrates partly because the resonant cavity is not tuned
for this particular frequency, and partly because the effective superstrate aperture is smaller than at 11.9
GHz. We see from Fig.~\ref{c1} that the plane-wave reflection coefficient of the 3-grid superstrate is still
very small at 10.4 GHz due to the dual resonance behavior. The corresponding effect is possibly seen as a much
more narrow beamwidth of the main lobe (in boresight direction) as compared with the 2-grid structure. However,
the finite structural dimensions clearly cause diffraction in the backward and sidelobe directions, and
therefore the obtained maximum directivity is only moderate. We believe, however, that after optimizing the
structure at 10.4 GHz the directivity could be enhanced (if desired antenna performance would be required over a
wider frequency band). At 10.4 GHz and also at 13.4 GHz the input resistance is visibly smaller than at 11.9
GHz, and this might lead to difficulties in achieving high radiation efficiencies when realistic losses are
taken into account (especially with the 2-grid structure at 10.4 GHz). Additional simulations (not shown) on the
current distribution induced to the wires in the 2-grid superstrate indicate that at 11.9 GHz there is only one
current maxima along the wires (in the $z$-direction, Fig.~\ref{antenna}), and the current amplitude in all the
wires (excluding the two pairs of wires closest to the dipole) is rather close to the same value. This means
that the superstrate aperture is rather uniformly excited, and a good boresight directivity is expected. At 10.4
GHz only every second pair is strongly excited. This is also the case at 13.4 GHz, in addition to this, at 13.4
GHz standing waves with three current maxima in $z$-direction are induced to the wires, and this is seen as the
beam splitting in the E-plane.

\renewcommand{\arraystretch}{1.5}
\begin{table}[t!]
\centering \caption{Simulated radiation characteristics.} \label{sim_table}
\begin{tabular}{|c|c|c|}
\hline \multicolumn{3}{|c|}{2 grids}\\ \hline
$f$ [GHz] & $Z_{\rm in}$ [Ohm] & $D^{\rm max}$ [dB] \\
\hline
11.9 & 50.2 + j86.8 & 15.3 \\
\hline
10.4 & 15.2 + j82.2 & 9.8 \\
\hline
13.4 & 33.3 + j96.2 & $\dag$ \\
\hline \hline \hline \multicolumn{3}{|c|}{3 grids}\\ \hline
$f$ [GHz] & $Z_{\rm in}$ [Ohm] & $D^{\rm max}$ [dB] \\
\hline
11.9 & 50.3 + j87.7 & 15.0 \\
\hline
10.4 & 33.0 + j96.5 & 9.0 \\
\hline 13.4 & 36.7 + j95.8 & $^\dag$\\ \hline
\end{tabular}

\smallskip $^{\dag}$ The maximum of directivity does not occur in the boresight direction.
\end{table}

\section{Conclusion}

We have presented a method to analyze the characteristics of a class of antenna superstrates based on periodical
grids of loaded and unloaded wires. In addition to this, the physics behind the radiation mechanism of such
superstrates has been clarified. The local field method has been used to solve the currents induced to the
superstrates by the incident plane-wave field, and the reflection and transmission properties of the
superstrates have been studied. The analytical results have been verified using numerical simulations. As a
practical application example we have considered finite-size superstrates illuminated by dipole antennas, and
discussed the effects of finite structural dimensions to the superstrate performance. A high-gain antenna
structure utilizing the proposed superstrates has been implemented in a commercial full-wave simulator, and the
results show promising antenna performance.

\section*{Acknowledgement}

One of the authors (P.M.T.~I.) wishes to thank Dr.~Pavel A.~Belov for useful discussions related to the wire
media modeling.

\end{document}